\documentclass[conference]{IEEEtran}
\IEEEoverridecommandlockouts
\usepackage{cite}
\usepackage{amsmath,amssymb,amsfonts,amsthm}
\usepackage{algorithmic}
\usepackage{graphicx}
\usepackage{hyperref}
\usepackage{textcomp}
\usepackage{xcolor}
\usepackage{xurl}

\newcommand{\Q}{\mathbb{Q}}

\newcommand{\vetx}{\boldsymbol{x}}
\newcommand{\vety}{\boldsymbol{y}}

\newcommand{\bb}{\begin{equation}}
\newcommand{\ee}{\end{equation}}
\newcommand{\bbb}{\begin{eqnarray}}
\newcommand{\eee}{\end{eqnarray}}
\newcommand{\benu}{\begin{enumerate}}
\newcommand{\eenu}{\end{enumerate}}

\newcommand{\bpm}{\begin{bmatrix}}
\newcommand{\epm}{\end{bmatrix}}

\newcommand{\one}[1]{\boldsymbol{1}_{#1}}

\newcommand{\cl}[1]{\mathcal{C}\ell_{#1}}

\newcommand{\basis}{\mathcal{E}=\{\boldsymbol{e}_0,\ldots,\boldsymbol{e}_{d-1}\}}

\def\BibTeX{{\rm B\kern-.05em{\sc i\kern-.025em b}\kern-.08em
    T\kern-.1667em\lower.7ex\hbox{E}\kern-.125emX}}

\newtheorem{definition}{Definition}

\newtheorem{remark}{Remark}
    
\begin{document}

\title{V-EfficientNets: Vector-Valued Efficiently Scaled Convolutional Neural Network Models
\thanks{Guilherme Vieira acknowledges the post-doctorate scholarship granted by Universidade Estadual de Campinas (UNICAMP). Marcos Eduardo Valle acknowledges financial support from the National Council for Scientific and Technological Development (CNPq), Brazil under grant no 315820/2021-7, the São Paulo Research Foundation (FAPESP), Brazil under grant no 2023/03368-0.}
}

\author{\IEEEauthorblockN{1\textsuperscript{st} Guilherme Vieira Neto}
\IEEEauthorblockA{\textit{Department of Applied Mathematics} \\
\textit{Universidade Estadual de Campinas (UNICAMP)}\\
Campinas, Brazil \\
vieirag@ime.unicamp.br}
\and
\IEEEauthorblockN{2\textsuperscript{nd} Marcos Eduardo Valle}
\IEEEauthorblockA{\textit{Department of Applied Mathematics} \\
\textit{Universidade Estadual de Campinas (UNICAMP)}\\
Campinas, Brazil \\
valle@ime.unicamp.br}
}
% \author{\IEEEauthorblockN{Anonymous Author}}

\maketitle

\begin{abstract}
EfficientNet models are convolutional neural networks optimized for parameter allocation by jointly balancing network width, depth, and resolution. Renowned for their exceptional accuracy, these models have become a standard for image classification tasks across diverse computer vision benchmarks. While traditional neural networks learn correlations between feature channels during training, vector-valued neural networks inherently treat multidimensional data as coherent entities, taking for granted the inter-channel relationships. This paper introduces vector-valued EfficientNets (V-EfficientNets), a novel extension of EfficientNet designed to process arbitrary vector-valued data. The proposed models are evaluated on a medical image classification task, achieving an average accuracy of 99.46\% on the ALL-IDB2 dataset for detecting acute lymphoblastic leukemia. V-EfficientNets demonstrate remarkable efficiency, significantly reducing parameters while outperforming state-of-the-art models, including the original EfficientNet. The source code is available at \url{https://github.com/mevalle/v-nets}.
\end{abstract}


\section{Introduction}

Over the past decade, convolutional neural networks (CNNs) have emerged as powerful tools for learning from spatial patterns in data. Their widespread adoption, coupled with advances in computational power, has propelled CNNs as the leading models for various computer vision tasks, particularly in image classification, object detection, and segmentation. During this time, numerous techniques and heuristics have been developed specifically for CNNs, while others have been adapted from fields such as optimization, mathematics, statistics, and physics \cite{Lecun2015DeepLearning,Gonzalez2018DeepNotes,li2021survey}.

% A defining strength of CNNs lies in their memory efficiency when processing images. 
Typical CNN architectures consist of convolution layers with multiple filters, alongside pooling layers, dense layers, and enhancements such as skip connections, normalization layers, and attention mechanisms \cite{He2015DeepRecognition,Hu2017Squeeze-and-ExcitationNetworks}. Performance in CNNs can often be improved by increasing the depth (number of layers), width (number of filters per layer), and resolution of input data. However, optimally allocating these additional resources is a non-trivial task. The class of EfficientNet models addresses this challenge by framing resource allocation as an optimization problem involving depth, width, and resolution \cite{tan2019efficientnet,tan2021efficientnetV2}. This design achieved state-of-the-art results in various image processing benchmarks, including ImageNet, Flowers, and CIFAR-100.

In parallel, vector-valued neural networks have gained significant attention in recent years. These models treat multiple feature channels as cohesive entities, representing and operating on data in higher-dimensional spaces \cite{Fan2020BackpropagationProducts,Valle2024UnderstandingProcessing}. This approach often leads to increased performance and reduced parameter counts, yielding more compact and accurate models \cite{Zhang2021BeyondParameters,Grassucci2022PHNNs:Convolutions}. Vector-valued neural networks also encompass hypercomplex-valued models, which leverage the geometric and algebraic properties of their underlying algebra \cite{Comminiello2024DemystifyingProcessing}. For instance, complex-valued neural networks are essential for effectively addressing phase-related information, such as that found in electromagnetism, light waves, quantum waves, and oscillatory phenomena \cite{aizenberg11book,Hirose2012Complex-valuedNetworks}. Quaternion-valued neural networks inherently account for spatial rotations and have promising applications in three-dimensional data modeling, as well as in processing polarization information widely used in sensing, imaging, and communication \cite{Parcollet2020ANetworks,Hirose2024QuaternionProcessing,Singh2024AScope,Chen2024RobustCommunications}. Furthermore, neural networks based on dual-quaternions exhibit rotational and translational equivariance properties, which are useful for describing or predicting spatial trajectories \cite{Vieira2023DualModelling}. Extending EfficientNet models to higher-dimensional spaces, therefore, combines the proven efficacy of this architecture with the representational efficiency of vector-valued neural networks, including hypercomplex-valued models.

This paper aims to introduce a class of vector-valued EfficientNet models. In addition to the convolutional and dense layers discussed in previous literature \cite{Fan2020BackpropagationProducts,Zhang2021BeyondParameters,Grassucci2022PHNNs:Convolutions}, we present vector-valued depthwise convolution layers. Similar to other layers, vector-valued depthwise convolution layers can be implemented using real-valued depthwise convolution available in deep learning libraries such as \texttt{PyTorch} and \texttt{TensorFlow}. However, unlike vector-valued dense and convolutional layers, they require additional attention to ensure consistency with the content of the vector entities. The organization of this paper is as follows: Section \ref{sec:maths} establishes the mathematical foundations of vector-valued operations. Section \ref{sec:layers} summarizes vector-valued layers used in the architecture and introduces vector-valued depthwise convolution. Section \ref{sec:effnet} discusses the core building blocks of EfficientNets and explains how to extend EfficientNets to higher-dimensional domains. Section \ref{sec:experimental} outlines the experiments conducted to detect acute lymphoblastic leukemia. Finally, Section \ref{sec:conclusion} offers concluding remarks.

\section{Mathematical Framework} \label{sec:maths}

Vector-valued neural networks are designed to treat multivariate information as single entities, considering the intercorrelation between features in advance. Therefore, an input image of width $W$, height $H$, and $d$ channels is interpreted as an array of size $W \times H$ of vectors in $\mathbb{R}^d$. For example, a color image can be viewed as an array of three-dimensional vectors representing the color values. Basic operations like addition and multiplication are performed on vectors in $\mathbb{R}^d$ instead of on scalars. These operations ensure that the $\mathbb{R}^d$ elements are treated as single entities. The following presents the basic mathematical framework for performing vector operations. 

\subsection{Definition and operations} \label{sub:def}

Non-associative algebras provide the mathematical background for vector-valued neural networks. The term ``non-associative'' means ``not necessarily associative''; it does not imply that associativity is not allowed:
\begin{definition}[Non-associative Algebra \cite{Schafer1961AnAlgebras}] \label{def:algebra}
A non-associative algebra $\mathbb{V}$ is a vector space over a field $\mathbb{F}$ equipped with a bilinear operation, called multiplication or product and denoted by the symbol ``$\times$'', which is not assumed to be associative. 
\end{definition}

In practice, we are mainly concerned with algebras over real numbers. Henceforth, we will consider only $\mathbb{F}=\mathbb{R}$. Furthermore, we focus only on finite-dimensional vector spaces. Precisely, we assume that $\mathbb{V}$ is a vector space of dimension $d$, that is, $\dim(\mathbb{V})=d$. 

Let $\mathcal{E} = \{\boldsymbol{e}_0,\boldsymbol{e}_1,\ldots,\boldsymbol{e}_{d-1}\}$ be an ordered basis for $\mathbb{V}$. Given $\vetx \in \mathbb{V}$, there exists a unique tuple $(\xi_0,\xi_1,\ldots,\xi_{d-1}) \in \mathbb{R}^d$ such that
\begin{equation}
\label{eq:sum}
\vetx = \sum_{i=0}^{d-1} \xi_i \boldsymbol{e}_i.
\end{equation}
The scalars $\xi_0,\xi_1,\ldots,\xi_{d-1}$ are the coordinates of $\vetx$ with respect to the basis $\basis$.

Being $\mathbb{V}$ a vector space, the sum and the multiplication by a scalar are defined as usual. Being a bilinear operation, the product of two vectors $\vetx = \sum_{i=0}^{d-1} \xi_i \boldsymbol{e}_i$ and $\vety=\sum_{j=0}^{d-1} \eta_j \boldsymbol{e}_j$ is characterized by the product of the basis elements $\boldsymbol{e}_i$ and $\boldsymbol{e}_j$ as follows:
\begin{equation}
    \label{eq:multiplication}
    \vetx \times \vety = \sum_{i=0}^{d-1} \sum_{j=0}^{d-1} \xi_i \eta_j (\boldsymbol{e}_i \times \boldsymbol{e}_j).
\end{equation}
However, the product of the basis elements $\boldsymbol{e}_i$ and $\boldsymbol{e}_j$ is also an element of $\mathbb{V}$. Thus, it can be written in terms of the ordered basis $\basis$ using the equation
\begin{equation}
    \label{eq:multiplication_table}
    \boldsymbol{e}_i \times \boldsymbol{e}_j = \sum_{k=0}^{d-1} \pi_{ijk} \boldsymbol{e}_k,
\end{equation}
where the terms $\pi_{ijk} \in \mathbb{R}$ represents the coordinate of $\boldsymbol{e}_i \times \boldsymbol{e}_j$ with respect to $\boldsymbol{e}_k$, for $i,j,k \in \{1,\ldots,d-1\}$. The products of the basis elements are often organized in a table, referred to as the multiplication table. \autoref{tab:multiplication-table} provides the multiplication tables of some notable four-dimensional non-associative algebras. 

\begin{table*}
\centering
    \caption{Multiplication tables of the quaternions ($\Q$), coquaternions ($\cl{1,1}$), tessarines ($\mathbb{T}$), and hyperbolic quaternions ($\mathbb{Y}$). \\ The bilateral identity for the product was suppressed (see Definition \ref{def:hypercomplex-algebra}).}
    \label{tab:multiplication-table}
\begin{tabular}{cccc}
{\small 
\begin{tabular}{c|rrr} 
             $\Q$ &  $\boldsymbol{e}_1$ & $\boldsymbol{e}_2$ & $\boldsymbol{e}_3$ \\ \hline
            $\boldsymbol{e}_1$ & $-\one{}$ & $\boldsymbol{e}_3$ & $-\boldsymbol{e}_2$ \\  
            $\boldsymbol{e}_2$ & $-\boldsymbol{e}_3$ & $-\one{}$ & $\boldsymbol{e}_1$ \\  
            $\boldsymbol{e}_3$ & $\boldsymbol{e}_2$ & $-\boldsymbol{e}_1$ & $-\one{}$\\  
        \end{tabular}} &
{\small 
\begin{tabular}{c|rrr}
     $\cl{1,1}$ & $\boldsymbol{e}_1$ & $\boldsymbol{e}_2$ & $\boldsymbol{e}_3$ \\ \hline
     $\boldsymbol{e}_1$ & $-\one{}$ & $\boldsymbol{e}_3$ & $-\boldsymbol{e}_2$ \\  
     $\boldsymbol{e}_2$ & $-\boldsymbol{e}_3$ & $\;\;\;\one{}$ & $-\boldsymbol{e}_1$ \\  
     $\boldsymbol{e}_3$ & $\boldsymbol{e}_2$ & $\boldsymbol{e}_1$ & $\one{}$\\  
    \end{tabular}   
    }
    &
    {\small 
    \begin{tabular}{c|rrr} 
             $\mathbb{T}$ & $\boldsymbol{e}_1$ & $\boldsymbol{e}_2$ & $\boldsymbol{e}_3$ \\ \hline
            $\boldsymbol{e}_1$ & $-\one{}$ & $\boldsymbol{e}_3$ & $-\boldsymbol{e}_2$ \\  
            $\boldsymbol{e}_2$ & $\boldsymbol{e}_3$ & $\one{}$ & $\boldsymbol{e}_1$ \\  
            $\boldsymbol{e}_3$ & $-\boldsymbol{e}_2$ & $\boldsymbol{e}_1$ & $-\one{}$\\  
        \end{tabular} 
        }
        & 
        {\small 
\begin{tabular}{c|rrr} 
             $\mathbb{Y}$  & $\boldsymbol{e}_1$ & $\boldsymbol{e}_2$ & $\boldsymbol{e}_3$ \\ \hline
            $\boldsymbol{e}_1$ & $\one{}$ & $\boldsymbol{e}_3$ & $-\boldsymbol{e}_2$ \\  
            $\boldsymbol{e}_2$ & $-\boldsymbol{e}_3$ & $\one{}$ & $\boldsymbol{e}_1$ \\  
            $\boldsymbol{e}_3$ & $\boldsymbol{e}_2$ & $-\boldsymbol{e}_1$ & $\one{}$\\  
        \end{tabular}
        }
\end{tabular}
\end{table*} 

\begin{remark}
The product of the basis elements completely characterizes the multiplication and, thus, the non-associative algebra. Hence, to determine a non-associative algebra, it suffices to provide a three-coordinate tensor $\boldsymbol{P}$ whose entries are $\pi_{ijk}$, for $i,j,k \in \{0,\ldots,d-1\}$.     
\end{remark}

Interestingly, the multiplication given by \eqref{eq:multiplication} can be computed efficiently using the matrix-vector product available in current scientific computing and deep learning libraries. To establish the relationship between vector multiplication and traditional matrix-vector product, let $\varphi:\mathbb{V} \to \mathbb{R}^d$ denote the isomorphism that maps $\vetx \in \mathbb{V}$ to its coordinates $(\xi_0,\ldots,\xi_{d-1}) \in \mathbb{R}^d$ with respect to the ordered basis $\basis$. The isomorphism $\varphi$, which depends on the ordered basis $\mathcal{E}$, is given by $\varphi(\vetx)=[\xi_0,\ldots,\xi_{d-1}]^T$, for all $\vetx \in \mathbb{V}$. Note that $\varphi$ is a linear mapping, that is, $\varphi(\vetx+\vety) = [\xi_0+\eta_0,\ldots,\xi_{d-1}+\eta_{d-1}]$.

Using the isomorphism $\varphi$ and recalling that the multiplication is a bilinear operation, we have
\begin{align*}
    & \varphi(\vetx\times \vety) = \varphi\left(\vetx \times \Big(\sum_{i=0}^{d-1} \eta_j \boldsymbol{e}_j \Big) \right) 
    = \sum_{j=0}^{d-1} \varphi(\vetx \times \boldsymbol{e}_j) \eta_j \\
    &= \begin{bmatrix}
& | &  \\
\ldots &\varphi(\vetx\times \boldsymbol{e}_j) & \ldots  \\
& | &  
\end{bmatrix} \begin{bmatrix}
    \eta_0 \\ \vdots \\ \eta_{d-1}
\end{bmatrix} \\
&= \begin{bmatrix}
& | &  \\
\ldots & \varphi\left(\sum_{i=0}^{d-1}\xi_i \boldsymbol{e}_i \times \boldsymbol{e}_j \right) & \ldots \\
& | &  
\end{bmatrix} \begin{bmatrix}
    \eta_0 \\ \vdots \\ \eta_{d-1}
\end{bmatrix} \\
&= \sum_{i=0}^{d-1}\xi_i \begin{bmatrix}
& | &  \\
\ldots & \varphi\left(\boldsymbol{e}_i \times \boldsymbol{e}_j \right) & \ldots \\
& | & 
\end{bmatrix} \begin{bmatrix}
    \eta_0 \\ \vdots \\ \eta_{d-1}
\end{bmatrix} \\
&= \mathcal{M}_L(\vetx) \varphi(\vety),
\end{align*}
where $\mathcal{M}_L:\mathbb{V} \to \mathbb{R}^{d \times d}$ is the mapping defined by  
\begin{equation}
    \label{eq:M_L}
    \mathcal{M}_L(\vetx) = \sum_{i=0}^{d-1}\xi_i \boldsymbol{P}_{i:}^T,
\end{equation}
with $\boldsymbol{P}_{i:}^T =  \begin{bmatrix}
\varphi(\boldsymbol{e}_i \times \boldsymbol{e}_0) & \ldots & \varphi(\boldsymbol{e}_i \times \boldsymbol{e}_{d-1}) \\
\end{bmatrix}$.
Equivalently,
\begin{equation}
\label{eq:PiT}
    \boldsymbol{P}_{i:}^T = \begin{bmatrix}
        \pi_{i00} & \pi_{i10} & \ldots & \pi_{i(d-1)0} \\
        \pi_{i01} & \pi_{i11} & \ldots & \pi_{i(d-1)1} \\
        \vdots & \vdots & \ddots & \vdots \\
        \pi_{i0(d-1)} & \pi_{i1(d-1)} & \ldots & \pi_{i(d-1)(d-1)} 
    \end{bmatrix}
\end{equation}
corresponds to the transpose of the matrix obtained by the $i$th slice of the tensor $\boldsymbol{P}$, whose entries are the parameters $\pi_{ijk}$ of the multiplication given by \eqref{eq:multiplication_table}. 

Concluding, the multiplication of $\vetx = \sum_{i=0}^{d-1} \xi_i \boldsymbol{e}_i \in \mathbb{V}$ and $\vety = \sum_{j=0}^{d-1}\eta_j \boldsymbol{e}_j  \in \mathbb{V}$ can be efficiently computed as follows using a matrix-vector product:
\begin{equation}
    \label{eq:practical-multiplication}
    \vetx\times \vety = \varphi^{-1}\left( \mathcal{M}_L(\vetx) \varphi(\vety) \right).
\end{equation}

\begin{remark}
    We would like to remark that \eqref{eq:practical-multiplication} results in the generalized multiplication defined by Zhang and collaborators \cite{Zhang2021BeyondParameters,Grassucci2022PHNNs:Convolutions}. Moreover, the multiplication of a non-associative algebra can be expressed using bilinear forms. Thus, it is also equivalent to the arbitrary bilinear product considered in  \cite{Fan2020BackpropagationProducts}.
\end{remark}

\subsection{Hypercomplex Algebras}

Hypercomplex algebras are significant non-associative algebras that encompass Cayley-Dickson and Clifford algebras, along with specific instances like complex numbers and quaternions \cite{Kantor1989HypercomplexAlgebras,Catoni2008TheSpace-Time}. Neural networks utilizing hypercomplex algebras have been successfully applied to a wide range of tasks in computer vision and signal processing \cite{Hirose2012Complex-valuedNetworks,Parcollet2020ANetworks,Hirose2024QuaternionProcessing,Comminiello2024DemystifyingProcessing}. Additionally, hypercomplex-valued neural networks demonstrate universal approximation capabilities on non-degenerate algebras \cite{Vital2022ExtendingNetworks,Valle2024UniversalNetworks}. A hypercomplex algebra is formally defined as follows:
\begin{definition}[Hypercomplex algebra \cite{Kantor1989HypercomplexAlgebras}]
    \label{def:hypercomplex-algebra}
A hypercomplex algebra, denoted by $\mathbb{H}$, is a finite-dimensional non-associative algebra over $\mathbb{R}$ with a bilateral identity for the product, often denoted by $\one{}$.
\end{definition}

Recall that a bilateral identity satisfies $\one{} \times \vetx = \vetx \times \one{} = \vetx$, for all $\vetx \in \mathbb{H}$. The identity $\one{}$ is usually the first element of an ordered basis of $\mathbb{H}$. Accordingly, the canonical basis for a hypercomplex algebra of dimension $d$ is $\tau = \{\one{},\boldsymbol{e}_1,\ldots,\boldsymbol{e}_{d-1}\}$. 

Quaternions, coquaternions, tessarines, and hyperbolic quaternions are four-dimensional hypercomplex algebras. We would like to recall that their multiplication tables with respect to the canonical basis $\tau=\{\one{},\boldsymbol{e}_1,\boldsymbol{e}_2,\boldsymbol{e}_3\}$ are provided in \autoref{tab:multiplication-table}. However, to fit the four multiplication tables within the page width, we suppressed the row and the column corresponding to the multiplication by $\one{}$, which is trivial.

The operations outlined in this section provide essential tools for building neural network models on vector-valued and hypercomplex-valued domains. They facilitate the efficient implementation of neural network layers, such as dense and convolution layers, within a vector-valued framework \cite{Zhang2021BeyondParameters,Grassucci2022PHNNs:Convolutions,Valle2024UnderstandingProcessing}. The following sections introduce the foundational concepts for implementing vector- and hypercomplex-valued EfficientNet models, including the vector-valued depthwise convolution layer.

\section{Basic Vector-Valued Layers} \label{sec:layers}

Vector-valued neural networks are obtained by replacing the arithmetic operations on real numbers with the corresponding non-associative algebra operations on vectors. Moreover, to take advantage of the current scientific computing and deep learning libraries, these operations are emulated using matrix and tensor operations \cite{Valle2024UnderstandingProcessing}. 

\subsection{Vector-Valued Dense and Convolutional Layers}

Zhang \textit{et al.} detailed in \cite{Zhang2021BeyondParameters} how vector-valued dense layers can be implemented efficiently using a traditional real-valued dense layer. Specifically, analogously to \eqref{eq:M_L}, the synaptic weight matrix $\boldsymbol{W}^{(\mathbb{R})}$ of the real-valued dense layer is given by the sum of the Kronecker product between the slices of the algebra tensor $\boldsymbol{P}_{i:}^T$ and the components $\boldsymbol{W}_i$ of the vector-valued synaptic weight matrix $\boldsymbol{W}^{(\mathbb{V})} = \sum_{i=0}^{d-1} \boldsymbol{W}_i \boldsymbol{e}_i$ as follows \cite{Valle2024UnderstandingProcessing}:
\begin{equation}
    \boldsymbol{W}^{(\mathbb{R})} = \sum_{i=0}^{d-1} \boldsymbol{W}_i \otimes \boldsymbol{P}_{i:}^T. 
\end{equation}
We would like to remark that the order of the terms in the Kronecker product must be considered depending on whether the input is multiplied from the left or the right when evaluating the matrix-vector product. Furthermore, while Zhang \textit{et al.}  \cite{Zhang2021BeyondParameters} discuss parameterized ``hypercomplex'' multiplication, their operation corresponds to the multiplication within an arbitrary non-associative algebra, as outlined in equation \eqref{eq:practical-multiplication}. Consequently, their implementation holds not only for hypercomplex-valued cases but also for vector-valued dense layers.

Grassucci \textit{et al.} later discussed the implementation of vector-valued convolution layers by utilizing traditional real-valued convolution layers \cite{Grassucci2022PHNNs:Convolutions}. Similar to the earlier work of Zhang \textit{et al.}, the filters from the traditional real-valued layer that emulates the vector-valued layer are derived using the Kronecker product of the vector-valued filter channels and slices of the algebraic tensor $\boldsymbol{P}$. As these layers are well-documented in the literature, let us turn our attention to the implementation of vector-valued depthwise convolution layer using the real-valued layers available in deep learning libraries.

\subsection{Vector-Valued Depthwise Convolution Layer}

\begin{figure}
    \centering
    \includegraphics[width=0.95\linewidth]{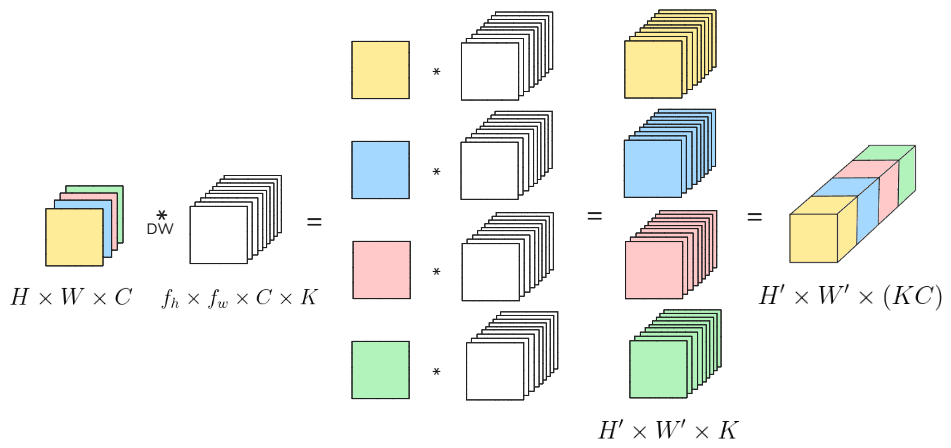}
    \caption{Traditional (real-valued) depthwise convolution.}
    \label{fig:real_depthwise}
\end{figure}

In regular convolution, each filter processes information from all input channels to produce a single-channel output. In contrast, a depthwise convolution layer applies filters individually to each input channel. The number of filters per channel is determined by a parameter known as the ``depth multiplier.'' Consequently, the number of output channels is a multiple of the input channels; specifically, it is equal to the depth multiplier times the number of input channels. In mathematical terms, let $\boldsymbol{I} \in \mathbb{R}^{H \times W \times C}$ be a real-valued image of height $H$, width $W$, and $C$ feature channels. The output $\boldsymbol{J} \in \mathbb{R}^{H' \times W' \times (KC)}$ of a traditional (real-valued) depthwise convolution by a bank of $K$ (depth multiplier) filters $\boldsymbol{F}\in \mathbb{R}^{f_h\times f_w \times C \times K}$ is $\boldsymbol{J} = \boldsymbol{I} \ast_{DW} \boldsymbol{F}$ given by the following equation for $c=0,\ldots,C-1$ and $k = 0,\ldots,K-1$:
\begin{equation}
\label{eq:Real-DWConv}
    \boldsymbol{J}_{:,:,(k+cK)} = \boldsymbol{I}_{:,:,c} \ast \boldsymbol{F}_{:,:,c,k}, 
\end{equation}
where $\ast$ represents a traditional 2D convolution. In other words, the $(k+cK)$th channel of $\boldsymbol{J}$ corresponds to the convolution of the $c$th channel of the image $\boldsymbol{I}$ by the $(c,k)$th filter $\boldsymbol{F}$. Fig. \ref{fig:real_depthwise} illustrates a traditional depthwise convolution operation. Finally, a depthwise convolution layer is defined by applying the activation function entry-wise to the sum of the bias term and the output of the depthwise convolution operation.  

The vector-valued depthwise convolution layer is analogous to a real-valued depthwise convolution layer but with vector-valued input, output, and filters. Moreover, traditional arithmetic operations are replaced by vector addition and multiplication from a non-associative algebra. Precisely, consider a vector-valued image $\boldsymbol{I}^{(\mathbb{V})} \in \mathbb{V}^{H \times W \times C}$, where $\mathbb{V}$ represents a $d$-dimensional non-associative algebra. This means the vector-valued image comprises $C$ feature channels, each with a dimension $d$. For example, when $d=3$, we can visualize $\boldsymbol{I}$ as a collection of $C$ color images concatenated together. Given a bank of $K$ vector-valued filters $\boldsymbol{F}^{(\mathbb{V})} \in \mathbb{V}^{f_h \times f_w \times C \times K}$, the output $\boldsymbol{J}^{(\mathbb{V})} = \boldsymbol{I}^{(\mathbb{V})} \ast_{DW} \boldsymbol{F}^{(\mathbb{V})} \in \mathbb{V}^{H' \times W' \times (KC)}$ from the vector-valued depthwise convolution operation is defined as
\begin{equation}
    \label{eq:V-DWConv}
    \boldsymbol{J}^{(\mathbb{V})}_{i,j,(k+cK)} = \sum_{m,n} \boldsymbol{I}^{(\mathbb{V})}_{(i-1)s_H+m,(j-1)s_W+n,c} \times \boldsymbol{F}^{(\mathbb{V})}_{m,n,c,k},
\end{equation}
where $s_H$ and $s_W$ denote the vertical and horizontal strides, respectively \cite{Geron19HandsOn}. We would like to remark that while the pixel indices are included, \eqref{eq:V-DWConv} is analogous to \eqref{eq:Real-DWConv}, differing only in that vectors replace the scalars and their corresponding operations are defined in the non-associative algebra $\mathbb{V}$. 

\begin{figure}
    \centering
    \includegraphics[width=0.95\linewidth]{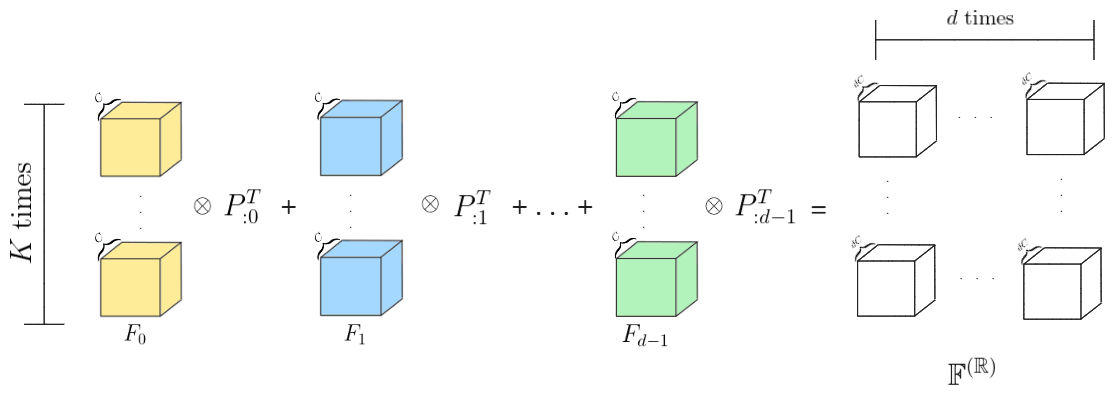}
    \caption{Computation of $\boldsymbol{F}^{(\mathbb{R})}$ by \eqref{eq:computing_F}.}
    \label{fig:computing_F}
\end{figure}

To leverage current deep learning libraries effectively, similar to the implementation of vector-valued dense and convolution layers \cite{Zhang2021BeyondParameters,Grassucci2022PHNNs:Convolutions,Valle2024UnderstandingProcessing}, the following demonstrates how to compute \eqref{eq:V-DWConv} using the real-valued depthwise convolution operation defined by \eqref{eq:Real-DWConv}. Given an ordered basis $\basis$ for $\mathbb{V}$, a vector-valued image $\boldsymbol{I}^{(\mathbb{V})} \in \mathbb{V}^{H \times W \times C}$ can be written as
$\boldsymbol{I}^{(\mathbb{V})} = \boldsymbol{I}_0 \boldsymbol{e}_0 + \ldots + \boldsymbol{I}_{d-1} \boldsymbol{e}_{d-1}$, where $\boldsymbol{I}_i \in \mathbb{R}^{H\times W\times C}$ is real-valued image for all $i=0,\ldots,d-1$. For practical aspects, the vector-valued image $\boldsymbol{I}^{(\mathbb{V})}$ 
can be identified with a real-valued image $\boldsymbol{I}^{(\mathbb{R})}$ obtained by concatenating channel-wise $\boldsymbol{I}_0,\ldots,\boldsymbol{I}_{d-1}$ as follows:
\begin{equation}
    \boldsymbol{I}^{(\mathbb{R})} = [\boldsymbol{I}_0,\ldots,\boldsymbol{I}_{d-1}]  \in \mathbb{R}^{H\times W \times (dC)}.
\end{equation}
Similarly, the vector-valued output $\boldsymbol{J}^{(\mathbb{V})} = \sum_{i=0}^{d-1} \boldsymbol{J}_i \boldsymbol{e}_i \in \mathbb{R}^{H'\times W' \times (KC)}$ can be identified with the real-valued image
\begin{equation}
    \boldsymbol{J}^{(\mathbb{R})} = [\boldsymbol{J}_0,\ldots,\boldsymbol{J}_{d-1}]  \in \mathbb{R}^{H'\times W' \times (dKC)}.
\end{equation}
% Conversely, a real-valued image with $dC$ feature channels can be interpreted as a vector-valued image with $C$ features of dimension $d$ each. 
Also, a bank of vector-valued filters can be written as $\boldsymbol{F}^{(\mathbb{V})} = \sum_{i=0}^{d-1} \boldsymbol{F}_i \boldsymbol{e}_i \in \mathbb{V}^{f_h \times f_w \times C \times K}$, where each $\boldsymbol{F}_i \in \mathbb{R}^{f_h \times f_w \times C \times K}$ is a bank of real-valued filters. Summing up the Kronecker product of $\boldsymbol{F}_i$, in the last two dimensions, and the slices of the algebra tensor $\boldsymbol{P}_{i:}^T$ yields 
\begin{equation}
\label{eq:computing_F}
    \boldsymbol{F}^{(\mathbb{R})} = \sum_{i=0}^{d-1} \boldsymbol{F}_{i} \otimes \boldsymbol{P}_{i:}^T \in \mathbb{R}^{f_h \times f_w \times (dC) \times (dK)}.
\end{equation}
Note that $\boldsymbol{F}^{(\mathbb{R})}$ represents a collection of $dK$ real-valued filters designed for images with $dC$ channels. Fig. \ref{fig:computing_F} illustrates how $\boldsymbol{F}^{(\mathbb{R})}$ is computed, with the filters arranged in a grid containing $K$ rows and $d$ columns.

\begin{figure}
    \centering
    \includegraphics[width=0.98\linewidth]{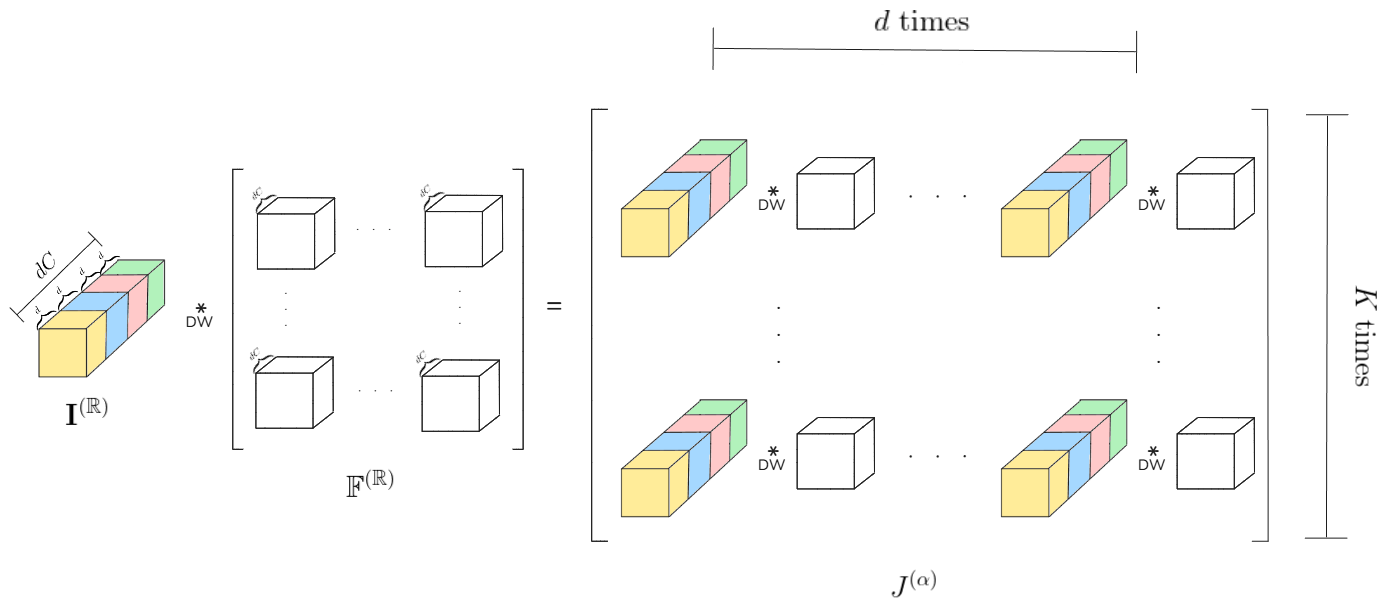}
    \caption{Augmented image $\boldsymbol{J}^{(a)} = \boldsymbol{I}^{(\mathbb{R})} \ast_{DW} \boldsymbol{F}^{(\mathbb{R})}$.}
    \label{fig:J_a}
\end{figure}

\begin{figure}[t]
    \centering
    \includegraphics[width=0.95\linewidth]{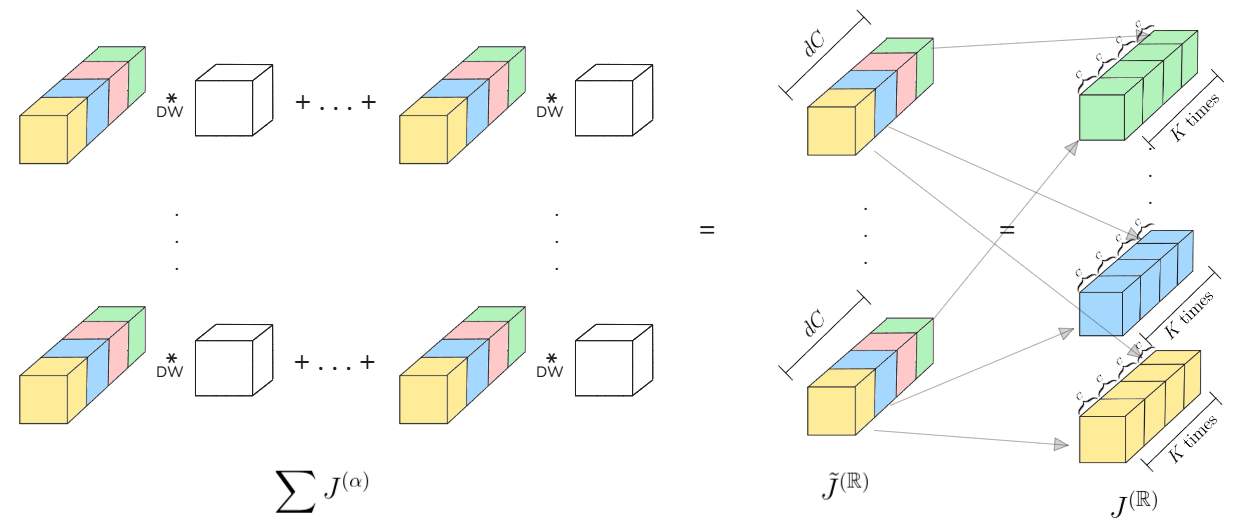}
    \caption{Computation of $\boldsymbol{J}^{(\mathbb{R})}$.}
    \label{fig:computing_J}
\end{figure}

The depthwise convolution operation performed on the real-valued image \(\boldsymbol{I}^{(\mathbb{R})}\) and \(\boldsymbol{F}^{(\mathbb{R})}\) results in an augmented real-valued image \(\boldsymbol{J}^{(a)}\) that has \(d^2 CK\) channels. Specifically, we have \(\boldsymbol{J}^{(a)} = \boldsymbol{I}^{(\mathbb{R})} \ast_{DW} \boldsymbol{F}^{(\mathbb{R})} \in \mathbb{R}^{H' \times W' \times (d^2 CK)}\). Fig. \ref{fig:J_a} depicts the computation of \(\boldsymbol{J}^{(a)}\). Similar to Fig. \ref{fig:computing_F}, the real-valued image \(\boldsymbol{J}^{(a)}\) is illustrated using a grid with \(K\) rows and \(d\) columns. 
Moreover, summing up appropriately (row-wise in the Fig. \ref{fig:J_a}) the channels of the augmented image $\boldsymbol{J}^{(a)}$, we obtain a real-valued image $\boldsymbol{\tilde{J}}^{(\mathbb{R})}$ which corresponds to the desired output except that the components of $\boldsymbol{J}^{(\mathbb{V})}$ are mixed. Therefore, the real-valued representation of $\boldsymbol{J}^{(\mathbb{R})}$ is obtained by permutating appropriately the channels of $\boldsymbol{\tilde{J}}^{(\mathbb{R})}$, as depicted in Fig. \ref{fig:computing_J}.
The source codes for implementing the vector-valued dense, convolution, and depthwise convolution layers are available on GitHub: \url{https://github.com/mevalle/v-nets}.

\section{Vector-Valued EfficientNet} \label{sec:effnet}

Balancing performance with resource constraints is a common challenge when designing deep convolutional neural networks. To tackle this issue, one effective strategy is to enhance a predefined network architecture to boost performance while staying within resource limitations. This enhancement can be achieved by adding more layers, increasing the number of filters in each layer, or increasing the resolution of the input.

The EfficientNet models, proposed by Tan and Le, achieved outstanding results on several benchmark tasks, outperforming state-of-the-art models with a significantly lower resource budget \cite{tan2019efficientnet,tan2021efficientnetV2}. They are powerful deep convolutional neural network models obtained through uniform scaling of the network parameters. The width, depth, and resolution of the networks are combined into a compound coefficient, which is used in an optimization problem for resource allocation. Below, we briefly review the key concepts of the EfficientNet models. The class of vector-valued EfficientNet (V-EfficientNets) is discussed in the sequel. 

\subsection{EfficientNet Models}\label{sec:realeffnet}

The first EfficientNets, referred here to as EfficientNetV1 models, were proposed by Tan and Le at ICML 2019 \cite{tan2019efficientnet}. The EfficientNetV1 family starts with the baseline network B0, designed by a neural architecture search to achieve an excellent trade-off between performance and resources. The other networks in the family, the EfficientNetV1 B1 to B7, are obtained by a clever compound scaling strategy.

\begin{figure}
\centering
\begin{tabular}{cc}
\parbox[t]{0.5\columnwidth}{\centering
a) Mobile inverted bottleneck (MBConv).} &
\parbox[t]{0.45\columnwidth}{b) Fused-MBConv.}\\
  \includegraphics[width=0.4\columnwidth]{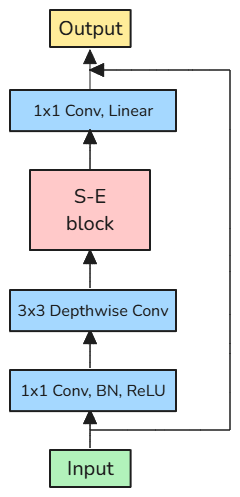} 
 &
  \includegraphics[width=0.4\columnwidth]{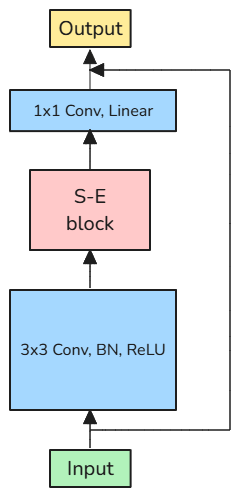}
\end{tabular}
% \caption{Building blocks of the EfficientNet-B0 model, where BN indicates batch normalization and ReLU is the rectified linear unit activation function.}
\caption{Building blocks of the EfficientNet models. Here, BN indicates batch normalization, and ReLU is the rectified linear unit activation function. EfficientNetV1 models use only the MBConv block, while V2 uses a combination of both blocks.}
\label{fig:EfficientNet-blocks}
\end{figure}

The main building block of the EfficientNetV1 models is the mobile inverted bottleneck (MBConv) block represented in Fig. \ref{fig:EfficientNet-blocks}a). The MBConv, which includes a depthwise convolution operation, was initially proposed as part of the MobileNetV2 architecture \cite{sandler2018mobilenetv2} and also features a squeeze-and-excitation block \cite{Hu2017Squeeze-and-ExcitationNetworks}.

Tan and Le have improved the EfficientNetV1 models at ICML 2021 \cite{tan2021efficientnetV2}. The improved EfficientNet models, titled EfficientNetV2, have a faster training speed and better parameter efficiency than the previous models. The baseline EfficientNetV2 model has also been designed using neural architecture search, but jointly optimizing performance, training speed, and parameter size. Moreover, besides the MBConv block, the search space has been enriched with additional operations, including the Fused-MBConv block. The Fused-MBConv block, proposed by Gupta and Tan \cite{Gupta2019efficientNetAutoML}, replaces the depthwise and expansion convolutions of the original MBConv block with a single regular convolution layer, as depicted in Fig. \ref{fig:EfficientNet-blocks}b). As noted by Tan and Le, depthwise convolutions tend to be resource-intensive in the early layers of the network but are highly effective in the later stages. Therefore, the EfficientNetV2 models utilize Fused-MBConv blocks in the early layers and regular MBConv blocks in the later layers.

% However, as pointed out by Tan and Le, it is not beneficial to replace all the MBConv blocks with Fused-MBConv, and the EfficientNetV2 models take into account an appropriate combination of these blocks \cite{tan2021efficientnetV2}.  

% \begin{figure}
%     \centering
%     \includegraphics[width=0.35\linewidth]{Figures/fused_mbconv.png}
%     \caption{Fused-MBConv block used in the EfficientNetV2 models.}
%     \label{fig:EfficientNetV2-blocks}
% \end{figure}

\subsection{V-EfficientNet: Vector-valued EfficientNet}

The vector-valued EfficientNet (V-EfficientNet) networks proposed in this work are based on the EfficientNetV2 models \cite{tan2021efficientnetV2}. These networks are created by replacing standard operations in real algebra with their corresponding vector-valued operations. This includes fundamental operations such as addition and multiplication, as well as the specific network layers described in Section \ref{sec:layers}. Indeed, vector-valued convolution and depthwise convolution layers are employed to construct the vector-valued equivalents of the MBConv and Fused-MBConv blocks (see Fig. \ref{fig:EfficientNet-blocks}), which are then arranged according to the EfficientNetV2 architecture.

Vector-valued networks (V-nets), which include hypercomplex-valued neural networks, are known for their parameter efficiency \cite{Vieira2024CliffordClassification,Grassucci2022PHNNs:Convolutions}. Unlike traditional real-valued neural networks, V-nets assume relationships among features in advance, allowing them to reuse parameters multiple times within a single feedforward step. They also take advantage of the algebraic and geometric properties inherent in the underlying non-associative algebra \cite{Comminiello2024DemystifyingProcessing}. Due to these characteristics, V-nets can be understood through the algebraic or geometric properties that yield physics-informed neural networks \cite{Comminiello2024DemystifyingProcessing,Hirose2024QuaternionProcessing}. Alternatively, they can be conceptualized regarding parameter reuse resulting from operations within non-associative algebras.

An important aspect of conceptualizing V-nets, particularly concerning feature relationships and parameter reuse, is their equivalence to traditional real-valued models. By extending from \(\mathbb{R}\) to a \(d\)-dimensional non-associative algebra \(\mathbb{V}\), we can substitute \(d\) real-valued parameters for a single vector in \(\mathbb{V}\). In the case of convolution or depthwise convolution layers, this creates an equivalence between a bank of filters \(\mathbf{F} \in \mathbb{R}^{f_h \times f_w \times C \times K}\) and \(\mathbf{F}^{(\mathbb{V})} \in \mathbb{V}^{f_h \times f_w \times C \times \frac{K}{d}}\), resulting in a reduction in parameters by a factor of \(1/d\) \cite{Grassucci2022PHNNs:Convolutions}. 
Alternatively, we can replace each real-valued filter with a vector-valued one, increasing the number of parameters by a factor of \(d\). For instance, the bank of real-valued filters \(\mathbf{F} \in \mathbb{R}^{f_h \times f_w \times C \times K}\) is simply substituted by a bank of vector-valued filters \(\mathbf{F}^{(\mathbb{V})} \in \mathbb{V}^{f_h \times f_w \times C \times K}\) in the convolution or depthwise convolution layers.

The cases in the previous paragraph represent two extremes of a spectrum concerning the number of parameters of a vector-valued network derived from a real-valued one: one involves compression by a factor of $1/d$, while the other entails upscaling by a factor of $d$. In the design of a V-EfficientNet model, we include a hyperparameter $\lambda \in \left[\frac{1}{d}, 1 \right]$, referred to as the \textit{architecture vectorization factor}. This factor represents the ratio of the number of features in the V-net to those in an equivalent real-valued network. On the one hand, \( \lambda = 1/d \) means that one vector comprises \( d \) real-valued features. Hence, a vector-valued convolutional layer is designed with a fraction of $1/d$ of the number of filters of the real-valued model, reducing the total number of parameters by a factor of $1/d$. On the other hand, one real-valued feature is mapped into a vector when $\lambda=1$. As a result, the number of filters in a vector-valued convolution layer is kept the same as that in a real-valued layer. This case leads to an increase in the number of parameters by a factor of $d$.

\section{Experimental Results} \label{sec:experimental}

While deep learning models have been successful in various computer vision tasks, they face significant challenges in the field of computer-aided medical diagnosis \cite{Zhou2021APromises,Rayed2024DeepChallenges}. One major issue is the scarcity of annotated medical images, which can hinder the training of these models. In this context, an effective model must be able to learn efficiently from limited data while minimizing the risk of overfitting. On one hand, EfficientNets have emerged as notable candidates for deep learning models due to their optimized allocation of parameters. Their successful applications in medical image classification have been demonstrated in studies such as \cite{Muduli2025DeepTechniques,Muhammad2025ALLApproach,Sousa2025EnsembleDiagnosis}, particularly in the context of leukemia diagnosis. On the other hand, hypercomplex-valued and, more generally, vector-valued deep learning models can leverage training data more effectively, resulting in improved performance compared to traditional real-valued models \cite{valous2025novelalgebras,Grassucci2022PHNNs:Convolutions,Comminiello2024DemystifyingProcessing,Vieira2022AcuteNetworks,Vieira2024CliffordClassification}. 
In light of these observations, this section presents the performance of the proposed V-EfficientNet in a classification task using the ALL-IDB2 database \cite{Labati2011All-IDB:Processing}.

Acute lymphoblastic leukemia (ALL) is a type of blood-related cancer that affects the bone marrow, leading to a high count of lymphoblasts in the peripheral blood. Diagnosis of ALL is usually performed manually by specialists based on visual analysis of blood smear images by counting the number of lymphocytes and lymphoblasts. Computer-aided diagnosis of ALL consists of performing a binary classification task on such images, and has proven to achieve high accuracy using a wide variety of deep learning techniques \cite{kumar2024integrating,Genovese2023DL4ALL:Detection,Genovese2022ALLNet:Networks,Vieira2022AcuteNetworks}. To the best of our knowledge, the highest reported accuracy is $98.46\%$ in the ALL-IDB2 dataset \cite{Genovese2022ALLNet:Networks}. The ALL-IDB2 is a public and free image-processing dataset containing 260 RGB-encoded images, each containing a single blood cell. This dataset was created by applying segmentation techniques to the original ALL-IDB dataset \cite{Labati2011All-IDB:Processing}. For the sake of comparison, we also conducted experiments using the ALL-IDB2. 

For this task, we build our V-EfficientNet models based on the EfficientNetV2-B0 \cite{tan2021efficientnetV2}, using the framework described in Sections \ref{sec:layers} and \ref{sec:effnet}. Based on previous works from the literature \cite{Vieira2022AcuteNetworks,Vieira2024CliffordClassification,Grassucci2022PHNNs:Convolutions}, we considered four-dimensional hypercomplex algebras to encode color information. Furthermore, we considered the smallest V-EfficientNetV2-B0 model by taking the architecture vectorization factor $\lambda=1/4$. Five network models were used: the baseline real-valued EfficientNetV2-B0 and four versions of our proposed V-EfficientNet (V2-B0). Each vector-valued model is based on a different four-dimensional algebra represented in Table \ref{tab:multiplication-table}: quaternions, coquaternions, tessarines, and hyperbolic quaternions \cite{Vieira2022AcuteNetworks}. Hyperbolic quaternions are a non-associative, non-commutative, non-degenerate algebra \cite{Takahashi2021ComparisonControl}. 

All networks are based on the EfficientNetV2-B0 structure and feature a real-valued dense output layer with two neurons. In the case of the vector-valued model, a real-valued dense layer corresponds to selecting one of the vector components, such as the real part of a hypercomplex-valued dense layer, as the model's output \cite{Valle2024UnderstandingProcessing}. The real-valued model has 5.86 million trainable parameters, while the vector-valued models consist of 1.72 million trainable parameters. This represents a reduction of approximately 70\% in the total number of parameters. Notably, this reduction is roughly equal to 3/4, which aligns with the vectorization factor of the architecture being reduced to 1/4 of the original network size.

To compare our results to previous approaches, we resized images to $128\times 128$. Furthermore, we performed a 50-50 train-test split as reported in  \cite{Genovese2021HistopathologicalDetection}. As in \cite{Vieira2022AcuteNetworks}, we carried out the experiments on RGB- and HSV-encoded images. Moreover, to mitigate chances of overfitting we applied data augmentation to the training images, comprising of a random rotation of magnitude up to $0.3\pi$, random translation of up to $10\%$ of image size horizontally and vertically, and a random (horizontal or vertical) flip.

The training was carried out using categorical cross-entropy loss, RMSProp optimizer with a learning rate of $0.001$, momentum of $0.9$, and weight decay of $5 \times 10^{-5}$. Based on \cite{tan2019efficientnet}, we also applied a learning rate decay schedule by $0.96$ every 2.4 epochs. Networks were trained for 400 epochs, and the experiment was performed 10 times per network. 

\begin{table}
    \centering
    \caption{Test set accuracy, mean $\pm$ standard deviation.}
    \begin{tabular}{|c|c|c|}  \hline
         \textbf{EfficientNetV2-B0} & \textbf{Color encoding} & \textbf{Accuracy}  \\ \hline 
         Real-valued & RGB & $98.23$ \scriptsize{$\pm 1.09$} \\
         Quaternion & RGB & $98.38$ \scriptsize{$\pm 0.80$} \\
         Quaternion & HSV & $\mathbf{99.23}$ \scriptsize{$\mathbf{\pm 0.91}$}\\
         Coquaternion & HSV & $98.08$ \scriptsize{$\pm 2.07$}\\
         Tessarines & HSV & $98.23$ \scriptsize{$\pm 1.50$}\\
         Hyperbolic quaternion & HSV & $\mathbf{99.46}$ \scriptsize{$\mathbf{\pm 0.60}$} \\ \hline
    \end{tabular}
    \label{tab:results}
\end{table}

Table \ref{tab:results} displays the average accuracy achieved by each model. First, note that the real-valued EfficientNetV2-B0 delivered good results, although its accuracy was slightly lower than that of the state-of-the-art models \cite{kumar2024integrating, Genovese2023DL4ALL:Detection, Genovese2022ALLNet:Networks}. This finding aligns with the notable performance reported for this model in the literature. Additionally, since the baseline version of the B0 model performed well for this image classification task, we refrained from considering the upscaled models B1-B7, as they contain more parameters. Among the proposed V-EfficientNet models, the quaternion-valued network (HSV) significantly outperformed the real-valued model. Furthermore, it surpassed the previously reported state-of-the-art performance, achieving a mean accuracy of 99.23\%. Finally, the V-EfficientNet model based on hyperbolic quaternions reached the highest accuracy of 99.46\%, outperforming the other deep networks by a solid margin and displaying a lower standard deviation. This result demonstrates both precision and reliability, representing the highest mean accuracy reported for this task in the literature to date.

\section{Concluding Remarks} \label{sec:conclusion}
Over the years, advancements in convolutional neural networks have mainly concentrated on improving performance, reducing resource demands, or both. EfficientNets are a family of models designed to balance performance, speed, and computational resources, resulting in compact networks that yield outstanding results. In this work, we extend this family of models into the vector-valued domain, further enhancing these features. Vector-valued networks are designed to treat multiple information channels as single entities, utilizing their intercorrelation to achieve superior performance. While previous works have addressed the implementation of vector-valued dense and convolutional layers \cite{Zhang2021BeyondParameters,Grassucci2022PHNNs:Convolutions,Valle2024UnderstandingProcessing}, this study is, to our knowledge, the first to define and utilize vector-valued depthwise convolution.
Additionally, we introduce a vectorization factor in our V-EfficientNet architecture, which acts as an additional scaling factor. This factor offers significant flexibility in network size while maintaining the proven structure of EfficientNet.

To evaluate our proposed family of models, we conducted a classification task using the ALL-IDB2 dataset. Our experiment resulted in V-EfficientNets-B0 with approximately 70\% fewer total parameters than the original EfficientNetV2-B0. Regarding accuracy, some of the vector-valued models matched the real-valued counterpart, while two V-EfficientNets significantly outperformed them. Precisely, the V-EfficientNet based on hyperbolic quaternions achieved remarkable results with an average accuracy of 99.46\%, the highest reported in the literature to date.

Finally, we note a performance gap between models that is potentially related to the algebraic and geometric properties of the underlying algebra. Choosing the best-suited algebra for a given task is a topic for future research.  

\bibliographystyle{IEEEtran}
\bibliography{references}

\end{document}